# MULTI AND INDEPENDENT BLOCK APPROACH IN PUBLIC CLUSTER

Z. Akbar and L.T. Handoko
Group for Theoretical and Computational Physic, Research Center for Physics, Indonesian Institute of Sciences, Kompleks Puspiptek Serpong, Tangerang 15310, Indonesia
zaenal@teori.fisika.lipi.go.id

## ABSTRACT

We present extended multi block approach in the LIPI Public Cluster. The multi block approach enables a cluster to be divided into several independent blocks which run jobs owned by different users simultaneously. Previously, we have maintained the blocks using single master node for all blocks due to efficiency and resource limitations. Following recent advancements and expansion of node's number, we have modified the multi block approach with multiple master nodes, each of them is responsible for a single block. We argue that this approach improves the overall performance significantly, for especially data intensive computational works.

## 1   INTRODUCTION

LIPI Public Cluster (LPC) is globally a unique infrastructure due to its openness [1,2,3]. This nature leads to some innovations on cluster architecture, especially so called multi block approach to enable multiple blocks of small cluster running simultaneously without any interruption among each other [4].

In multi block approach, all running blocks have a common single master node as shown in Fig. 1. This is actually motivated by resource (especially hardware) limitations. For instance all nodes were not equipped with storage media. So, the initial runtime environment contains several daemons called MPD's in each block were booted disklessly through network using embedded boot ROM in network cards attached in each node. The master node further works as a gateway for users, and all blocks have only one MPD in it. Therefore this master node is the last point for users accessing the cluster. Some benefits in this approach are :

- It avoids possible overlapping or interruption among the nodes owned by different users.
- Number of nodes in an allocated block can be changed easily.
- It prevents anonymous accesses to another blocks owned by another users.
- Very efficient in initial construction and further maintenance works.

As argued in previous paper [4], this technique is quite reliable and the overall performances are affected slightly.

However, we concern that the result is valid as long as the node's number is small, namely at the order of few nodes. Moreover, the approach is suitable for some computational works that are processor (including memory) intensive, but not for the others which are data intensive. In some processor intensive jobs, the data traffic through network among the nodes during computation period is relatively small. Because once all sub-jobs predefined in a parallel programming were sent to and initiated at the allocated nodes, each of them is executed independently in a node almost without any communication with the others.

In contrary, the case in some data intensive jobs like image mapping are quite different. This type of computational works usually require very intensive data exchange among the nodes during computation period. It is clear that then performance of cluster with conventional multi block approach would be decreased drastically in this case, since the master node is soon overloaded.

In LPC currently the allocated nodes in a block is usually few, and the data intensive job is extremely rare. Because the facility is moreless used for educational and training field for beginners in parallel programming. However, we anticipate further advancements of our users and the increasing level of their jobs in the near future. Also, recently we have upgraded the hardware environment to be more sophisticated, i.e. all nodes are currently equipped with storage medias. So, regardless with some benefits in the conventional multi block approach, we are now involved in expanding the multi block approach with an independent master node in each block.

In this paper, we first discuss the new approach, followed with the analysis performance before ending with conclusion.

## 2   MULTI BLOCK CLUSTER WITH INDEPENDENT MASTER NODES

Now we are ready to discuss the extended multi block approach with an independent master node in each block. The most common architectures for cluster computer are the symmetric and asymmetric clusters [5]. In the symmetric cluster all nodes are treated equally and accessible by external users. In contrast, there is only one node that is accessible by external users and performs as mediator between users and the rest nodes in the asymmetric cluster. In this case, there should be a public interface for user and another private interfaces for nodes.

In the LPC with independent master nodes, we extend it by adding some dedicated servers for specific jobs beside computational jobs owned by users. The architecture is shown in Fig. 2. In this approach, previous master node is transformed to be the gateway server mainly for users' access through internet. We also split the IO related jobs off to be handled by different dedicated server. Both servers are connected to all nodes through independent networks between them, namely the service and IO channels. The service server serves the web, ssh, monitoring services, and also stores all common binary programmes including libraries and compilers. On the other hand, the IO server is used for computational data communication, and serves the users' home directories, NIS and NFS services. Therefore we have implemented the IO channel on a gigabit local area network, while the service channel sits on regular fast ethernet based network.

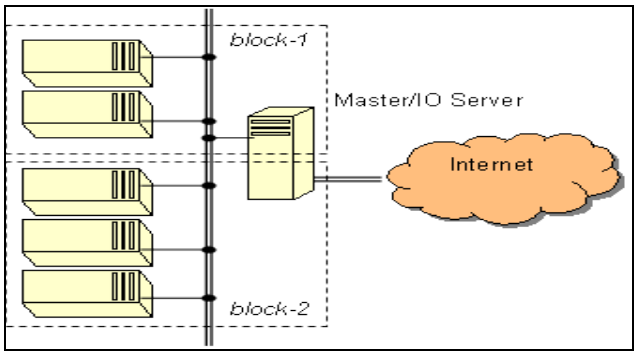

Figure 1. Multi block approach.

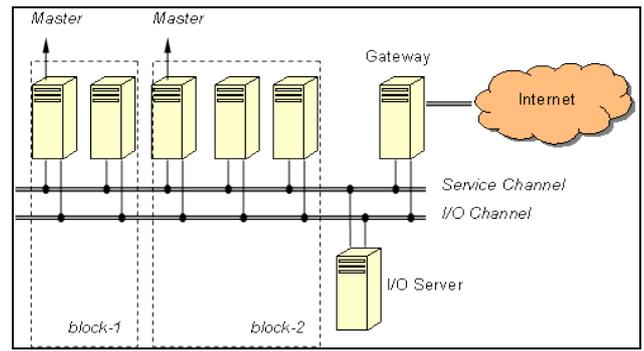

Figure 2. Multi and independent block approach.

This physical separation is urgent in the case of data intensive, or let us call it IO intensive, computational works as mentioned before. Many research works related to cluster's architectures are mainly intended for these IO intensive jobs. There are also another approaches to deal with, for example : making use of cache [6], or spreading of the IO processes from saturated nodes to other underloaded nodes [7].

Although the IO and service channels are separated physically, the blocks are separated logically as usual but with an independent master node in each of them. These master nodes function as interfaces between the gateway and the blocks. This means all commands sent by users over web in the gateway are forwarded to the master node of relevant block, and vice versa through the service channel.

Unfortunately, due to its nature in LPC there is no way to know in advanced the types of computational works being run by users, and actually we do not put any limitation on that. Therefore we are forced to accommodate both types easily and efficiently. Here efficient is in the sense that we should provide a block of cluster as close as to the user needs. For instance, IO intensive resources must not be provided to users who actually need only processor intensive ones. This is the main reason we still keep the conventional multi block approach in LPC. So during the initial resource allocation procedure done by administrator once a new registration has been processed [8], the administrator should determine the type of proposed computational works.

At time being, we have not yet implemented dynamical change of the cluster architecture partially, either conventional or independent multi block approaches. This means the whole LPC has used one of them, and not both of them at the same time.

Now let us proceed with the case that LPC deploys the independent multi block approach. All users' data are stored in the IO server, and then exported to all allocated nodes through Network File System (NFS) [9]. This solves the problem due to centralized data storage which could burdened the IO channel, and also enhances the flexibility of a block as well.

In the next section, we provide a performance analysis for typical IO intensive computational works.

## 3 PERFORMANCE ANALYSIS

In the present analysis, we are not using the same benchmarking programme mpptes as already done for the conventional multi block approach [4]. This programme has been developed by the Argonne National Laboratory [10], and runs under parallel programming environment Message Passing Interface (MPI) [11]. The reason is this time the analysis is focused on comparing the performances of the modified cluster architecture and the network being used. Instead we perform a ping-pong test based on the LAM-MPI [12].

Measurement is done by using twin independent blocks with exactly same specifications. Each consists of 4 nodes. We perform typical IO intensive computation by sending data with various size from 2 nodes in a block to the rest 2 nodes in the same block. First, this is executed only for one block, and further the same procedure is done for both blocks. Through this performance test, we could measure the reliability of independent multi block architecture against the growing number of blocks that could happen dynamically in the real daily operation.

First of all we compare the running time in fast-ethernet (FE) and gigabit networks (GE). The result for one block is depicted in Fig. 3 with time axis shows the round-trip time in microsecond. The message size is limited to around 33 Mb since the case in FE is not reliable anymore beyond it. In contrast with this, the GE is still reliable up to message size close to 1 Gb. Further we test for simultaneous twin blocks run in FE and GE as shown in Fig. 4. It is trivial that the performance in GE is significantly improved, and it is much better than in FE.

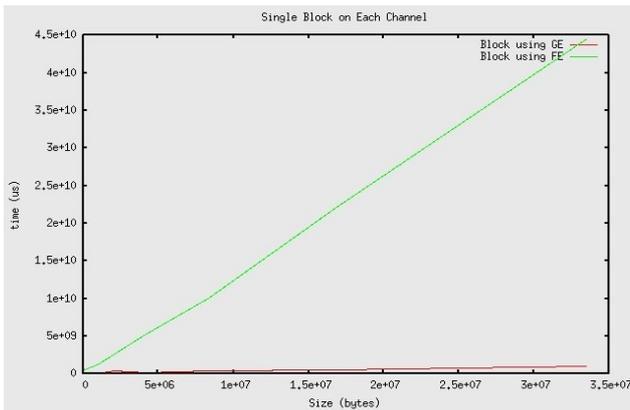

Figure 3. Comparison of round-trip times between FE (green line) and GE (red line).

## 4  CONCLUSION

These results prove our argument that, especially in multi-block cluster that is urgent for a public cluster, splitting off the networks and implementing gigabit network is very crucial to improve the whole performance.

We have shown that the independent multi block approach is much better especially for IO intensive computational works. This approach would complement the conventional multi block approach. Both of them should be combined together and implemented dynamically for node allocations according to the users' requests and types of parallel programmings being executed.

This study is very important for the administrator in accommodating the users' requests and allocating the resources for them. In the future we are going to develop simultaneous conventional multi block and independent multi block architectures in LPC to accommodate user's needs on IO intensive and processor intensive computational works at the same time.

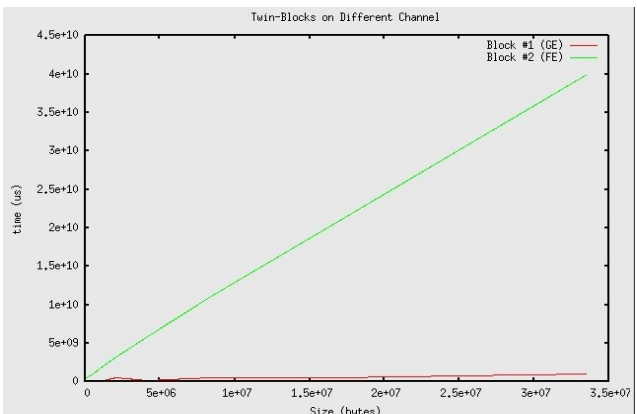

Figure 4. Round-trip times for twin blocks in FE (green line) and GE (red line).

## ACKNOWLEDGMENT

This work is financially supported by the Riset Kompetitif LIPI in fiscal year 2007 under Contract no. 11.04/SK/KPPI/II/2007 and the Indonesia Toray Science Foundation Research Grant 2007.